# Developer Prioritization in Bug Repositories

Jifeng Xuan
xuan@mail.dlut.edu.cn

He Jiang
jianghe@dlut.edu.cn

Zhilei Ren
ren@mail.dlut.edu.cn

Weiqin Zou
zou@mail.dlut.edu.cn

*School of Software, Dalian University of Technology*
*Dalian, China*

*Abstract*—Developers build all the software artifacts in development. Existing work has studied the social behavior in software repositories. In one of the most important software repositories, a bug repository, developers create and update bug reports to support software development and maintenance. However, no prior work has considered the priorities of developers in bug repositories. In this paper, we address the problem of the developer prioritization, which aims to rank the contributions of developers. We mainly explore two aspects, namely modeling the developer prioritization in a bug repository and assisting predictive tasks with our model. First, we model how to assign the priorities of developers based on a social network technique. Three problems are investigated, including the developer rankings in products, the evolution over time, and the tolerance of noisy comments. Second, we consider leveraging the developer prioritization to improve three predicted tasks in bug repositories, i.e., bug triage, severity identification, and reopened bug prediction. We empirically investigate the performance of our model and its applications in bug repositories of Eclipse and Mozilla. The results indicate that the developer prioritization can provide the knowledge of developer priorities to assist software tasks, especially the task of bug triage.

*Keywords*-developer prioritization; software evolution; bug triage; severity identification; reopened bug prediction

## I. INTRODUCTION

A bug repository is a vital database in modern software development. Many software projects create and maintain bug repositories for storing and updating the information of problems or suggestions about projects [1]. The widely available bug repositories have provided an important platform for investigating the quality of software [16]. With the growth in scale, developers in large projects must handle a large number of bugs in bug repositories. For example, from Oct. 2001 to Dec. 2010, the bug repository of an open source project, Eclipse [12], has recorded 333371 bugs, which are totally commented for 1544996 times by 34917 contributors around the world.

In *Peopleware*, Demarco & Lister [11] have proposed "*The major problems of our work are not so much technological as sociological in nature*". All the software artifacts in software repositories are created, updated, and studied by people. The social behavior of people has a significant impact on software development. Existing work has examined the social networks for some kinds of software repositories. For mailing list repositories, Bird et al. mine communication networks and discover the community structure from email archives [8]; Wolf et al. predict software build failures using social networks measures on the developer communication [34]. For change log repositories, Meneely et al. [25] and Pinzger et al. [27] build developer networks to predict software failures. For bug repositories, Hong et al. have divided the developer network into several communities, which identify the sub-groups of developer communication [17]. However, no prior work has considered the priorities of developers in bug repositories and its applications.

In this paper, we model the developer prioritization using a socio-technical approach to improve three predicting tasks centering on bug repositories. In contrast to dividing the developer network into communities in [17], we generate the developer prioritization by ranking all the participant developers of bug repositories. Based on our approach, we further study four Research Questions (RQs) with the experiments on two typical open source projects, namely Eclipse and Mozilla. We analyze the characteristics of developer prioritization to address the first three RQs, which study the developer priorities in products, the evolution over time, and the tolerance of noises, respectively; on the other hand, we address the last RQ by leveraging the developer prioritization to improve three typical tasks of bug repositories, including bug triage, severity identification, and reopened bug prediction. The experiments show that the developer prioritization is helpful to improve the predicting tasks in bug repositories. Especially, for bug triage, the average accuracy is improved up to 13% by combining the developer prioritization.

The primary contributions of this paper are as follows:

1. We identify the developer prioritization of bug repositories based on a socio-technical approach. In our work, each developer is mapped to a probability to indicate the priority in software development. To our knowledge, this is the first work for ranking developers with social networks in bug repositories.

2. We present detailed analysis of our developer prioritization. In the analysis, we examine the characteristics of the developer prioritization in bug repositories, including the developer priorities in products, the evolution, and the tolerance of noises.

3. We explore how to improve the tasks in bug repositories. We present the results of three typical tasks, i.e., improving bug triage by mixing the developer priorities, identifying bug severity by adding new features, and predicting reopened bugs by changing metrics. To our knowledge, this is the first work to evaluate the results of socio-techniques on bug-related tasks.

The remainder of this paper is organized as follows. Section II states the background. Section III shows the approach to recognize the developer prioritization. In Section IV, we study the characteristics of developer prioritization on bug repositories. In Section V, we describe how to improve software tasks by leveraging the developer prioritization. Section VI gives the threats to validity and Section VII shows the related work. Section VIII briefly summarizes this paper and presents the future work.

TABLE I   Two Bug Reports in Eclipse

| Commenter order | Bug ID 261871 | | Bug ID 264696 | |
|---|---|---|---|---|
| | Developer | Date & time | Developer | Date & time |
| Reporter | olivier_thomann | 09-01-21 12:39 | cwindatt | 09-02-12 10:02 |
| Commenter 1 | cwindatt | 09-01-21 12:54 | cwindatt | 09-03-12 16:25 |
| Commenter 2 | bcabe | 09-01-21 12:56 | ankur_sharma | 09-04-03 16:24 |
| Commenter 3 | olivier_thomann | 09-01-21 13:40 | cwindatt | 09-04-06 10:31 |
| Commenter 4 | bcabe | 09-01-21 14:08 | ankur_sharma | 09-04-14 17:01 |
| Commenter 5 | olivier_thomann | 09-01-21 14:18 | caniszczyk | 09-04-14 19:53 |
| Commenter 6 | bcabe | 09-01-21 15:43 | ankur_sharma | 09-04-15 06:49 |
| Commenter 7 | caniszczyk | 09-01-23 10:58 | cwindatt | 09-04-15 12:49 |

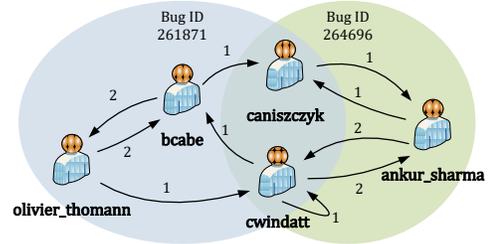

Figure 1. An illustration of the communication among developers. A link denotes a comment and the value of a link denotes the times of comments, e.g., the link with a value 1 from **bcabe** to **caniszczyk** means that **caniszczyk** has made one comment on **bcabe** (based on Commenters 6 and 7 in bug ID 261871).

## II. BACKGROUND

Bug repositories (also known as *bug tracking systems*) are deployed in software projects for the storage and management of bugs. A bug in bug repositories is recorded as a *bug report*, which is filled with the information of a software problem. Based on the bug reports, developers[1] can collect and reproduce bugs for bug fixing [40]. In empirical software engineering, many software tasks are investigated on bug repositories, e.g., assigning bugs to correct developers to reduce the time of bug fixing [1], [19], [32], summarizing a long bug report into a short abstract [29], detecting duplicate bugs to avoid repetitive operations [28], [35], [31], and characterizing factors of bug fixing [15].

In bug repositories, a bug report belongs to a component of a product. A product consists of multiple components and some products may share a common component. For example, Products **firefox** and **core** in Mozilla [22] use a common component, **file handling**. During the process of bug fixing, developers label a bug report with various statuses to denote a bug report as new, assigned, resolved, reopened, etc [18]. Once a bug report is created, any developer, who is interested in this bug, can make comments to communicate with relevant developers. In existing work on bug repositories, developers are always treated equally. However, the priority of a developer plays an important role in the tasks. For example, an active developer may make more contributions on bug fixing than an inactive one; an experienced tester may find bugs with higher severities than a common end user. In this paper, we rank all the developers of bug repositories to assist the tasks around bug repositories. We denote the process of generating the developer priorities as *developer prioritization*.

We take two bug reports (bug IDs 261871 and 264696) in Eclipse as examples to illustrate the developer prioritization. Both of these bugs are fixed bugs in Product **Plug-in Development Environment (PDE)**. In Table I, we list the details of these two bugs. Each bug report is reported by one developer and commented by developers for 7 times. We find that these two bugs share two common developers, caniszczyk and **cwindatt**. We illustrate the process of commenting among the developers in Fig. 1. An arrow with a value in Fig. 1 is called a *link*.

A question for us is how to prioritize the five developers for these two bug reports. This question is not easy to answer. Intuitively, both **caniszczyk** and **cwindatt** contribute to two bug reports while the other developers contribute to only one. On the other hand, **ankur_sharma**, **cwindatt**, and **bcabe** are very active since there are 3 links from each of them and 3 links to each of them (also a self-link for **cwindatt**). In this paper, we aim to model the developer prioritization. Furthermore, since developer factors are important for the metrics of software quality [40], another question is how to improve the software quality with the developer prioritization. In our work, we want to extract knowledge from the developer priorities to assist software development.

## III. IDENTIFYING THE DEVELOPER PRIORITIZATION

In this section, we model the developer prioritization by extending a social network technique. We present the framework of developer prioritization and propose four research questions for further studies.

### A. Framework of Developer Prioritization

Motivated by Fig. 1, we define the developer prioritization in our work.

**Developer prioritization** is a process to assign a priority to each developer in a bug repository and to rank all the contributions of developers to assist software tasks.

We extend a recent socio-technical approach proposed by Lü et al. [21] to identify the developer prioritization. In their work, a leadership network is proposed to investigate the social analysis between leaders and fans on an online bookmarking website. In our developer prioritization, we transfer their work to developers in a bug repository. We contribute in two extensions. First, we adapt the original binary weights of links to integer weights based on the number of comments.

---

[1] In this paper, "developers" refers to the people who contribute to a bug repository. We follow existing work [17], [19] to use the term developers in a broad sense, including reporters, programmers, testers and active end users. We denote a developer with the user name in the email instead of the real name.

**Algorithm 1**. Framework of Developer Prioritization

**Input:** developer $d_i$ ($1 \leq i \leq n$), links among $d_i$
**Output:** final score $S_i$ for each developer $d_i$
1   add a virtual developer $d_0$ and add bi-directional links with $d_0$;
2   set initial scores $s_i(0) = 1$ ($1 \leq i \leq n$) and $s_0(0) = 0$;
3   **for** $t = 1$ **to** $t_c$ **do**    // $t_c$ is the time for convergence
4     calculate the score of each $d_i$ ($0 \leq i \leq n$) at time $t$ with (1);
5   calculate the final score of each $d_i$ ($1 \leq i \leq n$) with (2).

Second, we propose a topic-based prioritization by specifying a product or a component.

We present the brief framework of the developer prioritization in Algorithm 1. Given $n$ developers in a bug repository, the goal of developer prioritization is to generate a score $S_i$ for each developer $d_i$ and rank all the developers based on these scores. A weight $w_{ij}$ denotes the number of all the comments in a link from a developer $d_i$ to $d_j$. If no link exists between $d_i$ and $d_j$, $w_{ij} = 0$. Specially, we remove all the self-links since we omit the influence of comments from a developer to himself/herself (e.g., the self-link of **cwindatt** in Fig. 1). To build a connected graph based on all the links, a virtual developer $d_0$ is added to connect all the developers. Then, we add a bi-directional link between each original developer $d_i$ and $d_0$. The weights of this link, $w_{i0}$ and $w_{0i}$, are set to 1 for each $d_i$ ($1 \leq i \leq n$). For each developer, let $o_i$ ($0 \leq i \leq n$) denote the out-degree of $d_i$. Note that both out-degrees and in-degrees are informative to study the social networks in bug repositories [17]. In our work, we use out-degrees to reflect the influences of both reporters and commenters.

We consider that the developer prioritization is generated based on the changes of time series. We denote $s_i(t)$ as the score of developer $d_i$ ($0 \leq i \leq n$) at time $t$. Thus, we calculate this score,

$$s_i(t) = \sum_{j=0}^{n} w_{ji} s_j(t-1) / o_j \quad (1)$$

The initial score $s_i(0) = 1$ for $1 \leq i \leq n$ and $s_0(0) = 0$. Lü et al. have proved that the above model can converge after finite time [21]. Given the convergence time $t_c$, we generate the final score of $d_i$ ($1 \leq i \leq n$),

$$S_i = (s_i(t_c) + s_0(t_c)/n)/M \quad (2)$$

where $M$ is a parameter for normalization and $M = s_0(t_c)/n + max_{1 \leq i \leq n} s_i(t_c)$. Thus, each developer in the bug repository is assigned with a score $S_i$ ($0 < S_i \leq 1$).

We rank all the participant developers by their scores in descending order. In other words, a developer in a top rank owns a higher priority than a developer in a bottom rank. We apply the developer prioritization to the example in Fig. 1. The scores of the five developers are $S_{ankur\_sharma} = 1.0$, $S_{cwindatt} = 0.9012$, $S_{bcabe} = 0.7956$, $S_{caniszczyk} = 0.7274$, and $S_{olivier\_thomann} = 0.6763$, respectively; the developer **ankur_sharma** has the highest priority among these five developers.

The scores of these five developers illustrate that Algorithm 1 can quantify the priorities of developers. In contrast to the same value by measuring out-degrees (e.g., the same out-degree of **ankur_sharma**, **cwindatt**, and **bcabe**), Algorithm 1 distinguishes the developers with the same out-degree.

TABLE II SCALES OF DATA SETS

| Projects | #Bug reports | #Developers | #Comments | #Products | #Components | Period |
|---|---|---|---|---|---|---|
| Eclipse | 332142 | 34917 | 1544996 | 160 | 835 | 01-10-10 to 10-12-31 |
| Mozilla | 599870 | 146500 | 4543146 | 58 | 782 | 98-04-07 to 10-12-31 |

Based on Algorithm 1, the developer prioritization can be adapted to a topic-based model. In a bug repository, we usually focus on a specified product or component of the project. A straightforward choice is to denote a topic with a product or a component. For example, given a product, we can generate the developer priorities based on the bug comments related to this product. Product-based or component-based developer prioritization is helpful to study the priorities of developers in a specified part of the project.

### B. Research Questions

We propose four Research Questions (RQs) to investigate the developer prioritization. These four RQs are divided into two categories, namely the characteristic analysis and the applications. We answer these two categories of RQs in Section IV (RQ1-RQ3) and Section V (RQ4), respectively.

**RQ1**. Does the developer prioritization for the whole project differ from the one for a product?

**RQ2**. How does the developer prioritization evolve over time?

**RQ3**. Is the developer prioritization tolerant to noisy comments?

To analyze the characteristics of the developer prioritization, in RQ1, we study the differences of the developer prioritization between the products and the whole project; in RQ2, we study the evolution of the developer prioritization over time; in RQ3, we examine the tolerance of noises for the developer prioritization.

**RQ4**. Can we use the developer prioritization to assist the existing tasks in bug repositories?

An important problem is to explore the applications of the developer prioritization. In RQ4, we investigate how to incorporate the developer prioritization to improve typical tasks in bug repositories.

## IV. ANALYZING THE DEVELOPER PRIORITIZATION

To explore the answers to the above four RQs, we conduct experiments on bug repositories of two open source projects, Eclipse and Mozilla. In this section, we present the details of the data collection and investigate the answers to RQ1 - RQ3.

### A. Data Collection

We analyze the characteristics of the developer prioritization based on bug repositories of Eclipse and Mozilla. These two projects have attracted wide interests since both of them are large scale and open source projects. In our work, we collect all the bug reports before 2011, i.e., bugs 1-333371 in Eclipse and bugs 1-662320 in Mozilla. The details of our data sets are listed in Table II. Note that the number of bug reports does not equal to the range of bug IDs since some bug reports are removed in development (e.g., bug 5315 in Eclipse) or not anonymously accessible (e.g., bug 400020 in Mozilla).

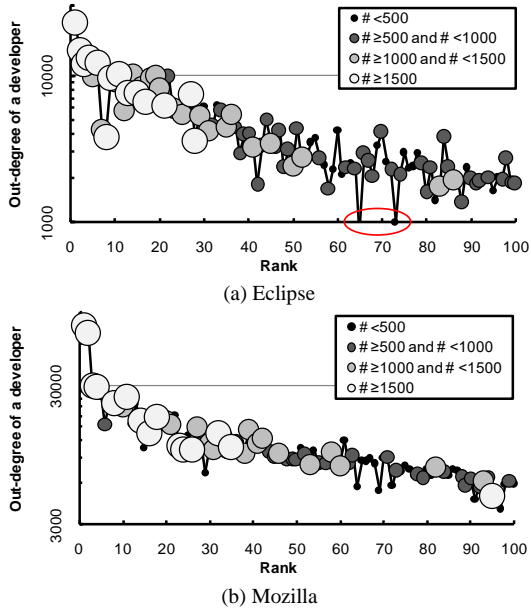

Figure 2. Out-degrees for top 100 developers. The diameter of a circle denotes the number of bug reports for a developer. For example, the smallest circles denote developers who have fixed less than 500 bugs. Note that the vertical axis is on a log scale.

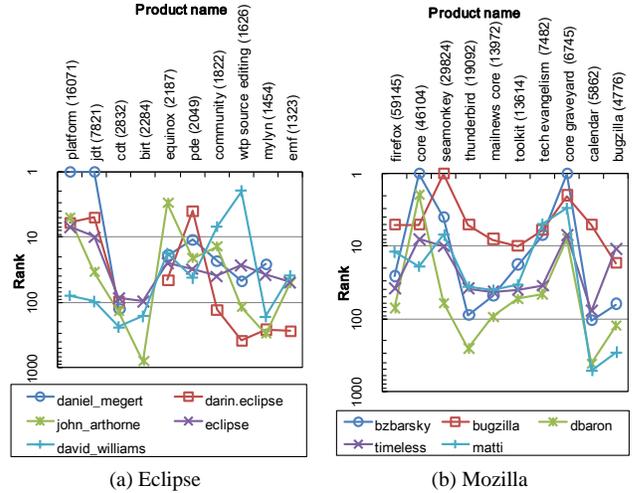

Figure 3. Ranks in 10 active products for top 5 developers in the whole project. For each project, the selected products are 10 products, which are contributed by most developers. In horizontal axis, the number following with a product name denotes the number of participant developers. Note that values in the vertical axis are in reverse order.

For each bug report, we extract the bug ID, the reporter, the fixer, the summary, the description, the creating time, and the comments. For each comment on a bug report, we extract the commenter and the commenting time. Since the fixer of a bug report is not always correctly labeled, the heuristics in [1] are used to recognize the correct fixers in bug repositories. In this section, we mainly focus on the comments of bug reports while in Section V, we will further study the texts of bug reports, e.g., the summary and the description.

### B. Answer to RQ1, Developer Prioritization in Products

Most open source projects consist of multiple products, each of which can be viewed as a sub-project for a set of individual requirements. In practice, a developer can participate in multiple products since the experience from one product may guide the development of another one. In this sub-section, we examine the changes of developer prioritization between the whole project and its products.

Before studying the developer prioritization for products, we first illustrate the developer rankings in the whole projects in Fig. 2. Three indicators of developers are used, namely the rank in the developer prioritization, the out-degree in a bug repository, and the number of fixed bugs. We choose the out-degree as an indicator since the developer prioritization is constructed based on the out-degrees of developers (in Algorithm 1). Note that since not all the bugs are fixed in bug repositories, we only count the number of fixed bugs based on bug reports with the resolution "fixed" to simplify the statistics.

In Fig. 2, a developer with a large out-degree leads to a high priority in developer rankings. Moreover, most of developers with high priorities have fixed a large amount of bugs, e.g., most of the largest circles for developers (who have fixed over 1500 bugs) lie in top 40 ranks. We can observe that the curve of Eclipse is not as stable as that of Mozilla. In Eclipse, most of the out-degrees are over 2000. In both Eclipse and Mozilla, there are some developers who are dominant in both the priority and the number of fixed bugs, i.e., top 2 developers in Eclipse and top 4 developers in Mozilla. These dominant developers may be the experienced experts in software development [2]. Note that some developers, who have only fixed a small number of bugs, also have high priorities, e.g., a developer in Mozilla is ranked in top 30 and has fixed less than 500 bugs. This fact is caused by the different duties of developers, e.g., an active developer may be not a fixer but a tester.

To observe the differences of developer prioritization between the whole project and its products, we show the ranks in 10 products for 5 developers in Fig. 3, who are ranked as the top 5 in the whole project (i.e., top 5 developers in Fig. 2). The curves for most developers have the similar trend, e.g., in Mozilla, the 5 developers have high priorities in Product **core** and have low priorities in Product **calendar**. An exception is the rank for Product **wtp source editing** in Eclipse, which widely distributes between 1 and 1000. Moreover, a top developer in the whole project may contribute little to a product or not participate in a product, e.g., in Eclipse, the developer **john_arthorne** has ranked around 900 in Product **birt** and the developer **darin.eclipse** has not contributed to **birt**.

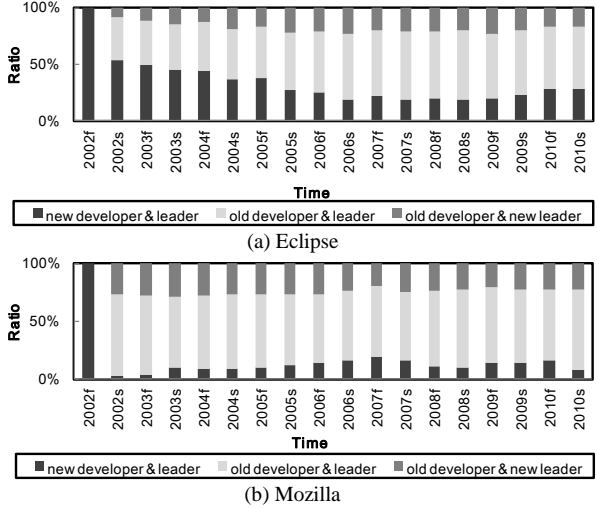

Figure 4. Percentage for developers and leaders over time. The leaders (top 1000 developers) in each unit time is considered. The ratio of a new developer who is a leader, the ratio of an old developer who is a leader, and the ratio of an old developer who is new as a leader are labeled in black, light gray, and dark gray, respectively.

**Answer to RQ1**. The developer prioritization in products differs from that in the whole project. Specially, certain top developers in the whole project may contribute little or nothing to a product. Among the participated products, top developers contribute much to most of products.

*C. Answer to RQ2, Evolution of Developer Prioritization*

For a large scale software product, personnel changes are common in development. An active developer may be inactive in a period, or even retire [10]. In this sub-section, we investigate the evolution of the developer prioritization over time.

Based on Table II, Eclipse has been developed for over 9 years and Mozilla has been for over 12 years. To investigate the evolution, we select bug reports in 9 continuous years from Jan. 2002 to Dec. 2010 (since Eclipse is started from Oct. 2001). We choose half a year as the unit time by following [17]. Thus, the time for each project is divided into 18 periods. We denote the first half of a year with "f" and the second half of a year with "s". For each unit time, we generate the developer prioritization in the whole projects and analyze the evolution of leading developers over time.

We illustrate the changes of top developers in Fig. 4. To simplify the expressions, we denote the developers who are ranked in top 1000 in the whole project as *leaders*. Given a period, a developer is called an old leader or a new leader, if he or she is a leader in the last period or not, respectively. As shown in Fig. 4, the ratios in Mozilla are more stable than those in Eclipse. Moreover, for both projects, the changes after 2005 are stable. In Eclipse, the ratio of new developers who are also leaders is over 20% in each unit time. From 2002s to 2006s, the ratio of old developers who are also old leaders increases with time. In Mozilla, the ratio of new developers who are leaders is less than 20%.

In both Eclipse and Mozilla, the developer prioritization changes over time. We list two possible reasons for this fact.

TABLE III RATIOS OF DEVELOPERS AND COMMENTS IN DATA SETS

| Project | Ratio | >15 comments | >10 comments | >5 comments | Original | >3 words | >6 words | >9 words |
|---|---|---|---|---|---|---|---|---|
| Eclipse | Developer (%) | 81.24 | 81.47 | 82.51 | 100.00 | 99.61 | 98.48 | 97.02 |
| | Comment (%) | 95.55 | 96.46 | 97.95 | 100.00 | 94.14 | 81.84 | 71.77 |
| Mozilla | Developer (%) | 79.42 | 79.67 | 80.81 | 100.00 | 99.80 | 99.28 | 98.38 |
| | Comment (%) | 94.77 | 95.70 | 97.29 | 100.00 | 95.92 | 88.37 | 80.78 |

One is that the developer prioritization always changes because of the complexity of such large projects; the other is the unit time in our experiments is not short enough to recognize the fixed developer prioritization. A further study is needed to explore the appropriate unit time to model unchanged developer prioritization in projects.

**Answer to RQ2**. The developer prioritization evolves over time. A new developer can join the projects and become a developer with high priorities.

*D. Answer to RQ3, Tolerance of Noisy Comments*

The process of fixing and localizing bugs suffers from the bad quality of bug reports [18]. Noises in bug reports are common in bug repertories. For example, in the bug report with ID 1 of Eclipse, Comment 45 is just a test of a user account, which has nothing to do with the content of the bug. Since the developer prioritization is built based on the bug comments, we investigate whether the developer prioritization is sensitive to noisy bug comments.

Due to the lack of existing method to identify noisy comments, we label noisy comments with a heuristic. We consider two types of comments as noises, namely the comments by inexperienced developers and the comments written in very few words. Note that not all the comments in these two types are noisy, e.g., Comment 2 of Bug 250031 in Eclipse, only containing a full stop, can be viewed as a noisy comment while Comment 2 of Bug 250033, only containing a word "verified", is a useful comment. In this paper, we directly treat the two types of bug reports as noisy comments without further identification. In Table III, we present the ratios of developers and comments for the data sets, which are generated by removing two types of comments. The seven columns denote three data sets by removing comments of inexperienced developers (5, 10, and 15), the original data set, and three data sets by removing comments written in few words (3, 6, and 9).

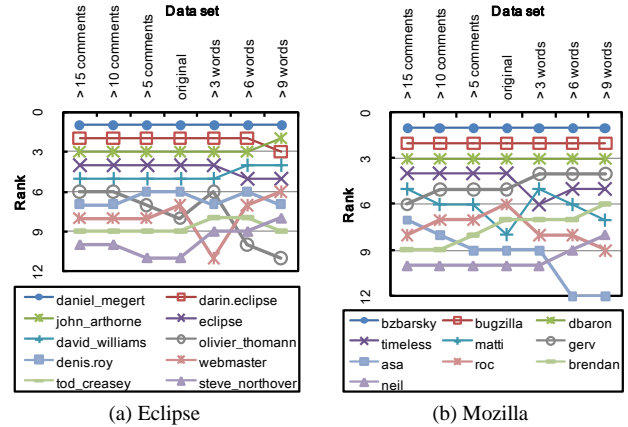

Figure 5. Changes for the ranks of top 10 developers among the original data set and new data sets after removing noises. The selected developers are top 10 developers in the original data set.

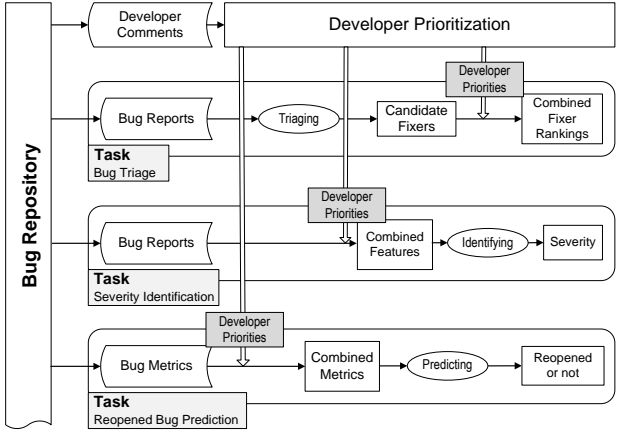

Figure 6. Leveraging the developer prioritization to predict tasks around bug repositories. We use the developer prioritization to assist three tasks. A circle denotes an action of predicting.

We illustrate the changes of the developer prioritization in Fig. 5. We can find that our approach for the developer prioritization is tolerant to noisy comments. The change among each data set is inconspicuous. This fact coincides with our expectation since we always focus on the developers with high priorities, who contribute many comments to bug repositories. The noisy comments removed in our experiments only affect the developers with few comments.

**Answer to RQ3**. The developer prioritization in our work is insensitive to noisy comments. Therefore, the developer prioritization can be used to handle the real data sets, which include noisy bug comments.

## V. LEVERAGING THE DEVELOPER PRIORITIZATION

In this section, we explore the results of leveraging the developer prioritization to assist the tasks in bug repositories. We answer RQ4 by examining the effects on three typical tasks, i.e., bug triage, severity identification, and reopened bug prediction. All these tasks have been addressed to improve the quality of software development. We select such three tasks since they cover various aspects of predictive tasks. In details, bug triage [1] is a multiple-class task based on bug repositories; severity identification [20] is a binary-class task based on bug repositories; and reopened bug prediction [30] is a binary-class task based on both bug repositories and change log repositories.

In this section, we combine the input or the output of a task with developer priorities obtained by the developer prioritization. By combining with the input, we add new features to the predictive model while by combining with the output, we update the results of a task. In Fig. 6, we briefly summarize the process for improving the results of the three tasks by combining the developer priorities.

### A. Bug Triage

Bug triage is a typical problem in software maintenance, which aims to predict a correct developer for a new-coming bug [1]. Traditionally, a human developer (also called *triager*) assigns new bugs to candidate developers. Automatic approaches for bug triage have been developed to reduce time and labor cost. Most of existing work models bug triage as text classification and improves the accuracy of bug triage based on the knowledge from bug repositories [1], [19], [38], [32].

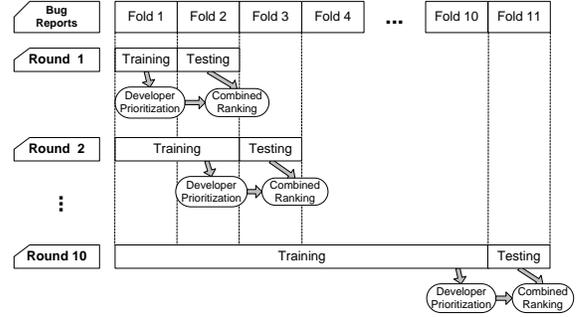

Figure 7. Bug triage combined with the developer prioritization based on the incremental learning. In Round $r$, the first $r$ folds are used for training a classifier and generating the developer prioritization while the $(r+1)$th fold is used for testing. The developer prioritization is combined with the result of testing to form a new developer list.

In this paper, we consider improving bug triage with the developer prioritization obtained from bug repositories. For a predicted list of developers by a classifier, we rank these developers by the priorities. Thus, the developer prioritization is used to discriminate the developers with similar probabilities in the prediction.

We evaluate the results of bug triage with the incremental learning framework, in which we chronologically sort all the bug reports and divide these bug reports into 11 folds [6]. Thus, we perform experiments in 10 rounds. In each round, we generate the developer prioritization from the training set and combine the developer priorities with the predicted results of the classifiers. In Fig. 7, we present the evaluation framework in our work.

We validate our approach on the bugs from 200001 to 300000 for Eclipse and bugs from 400001 to 500000 for Mozilla. We follow the existing work [1], [9] to remove the non-fixed bug reports (only bug reports with the resolution "fixed" are left) and inactive developers (in our work, developers who have fixed less than 50 bugs are removed). As a result, 49762 bug reports of Eclipse and 30609 bug reports of Mozilla are left as data sets. For each bug report, the title and the description are extracted as an input text while the developer who has fixed this bug is extracted as a label for the classifier. We convert the bug reports into vector space model by tokenizing the sentences into terms. We perform the techniques of removing stop words, stemming, and *tf-idf* (a weighted term-frequency approach [33]) to generate the final data sets. We evaluate the experiments with the accuracy of top-$k$ predicted developers since a recommendation list is always employed [2]. The accuracy is calculated as $Accuracy_k = \frac{\text{\# correctly predicted bugs}}{\text{\# all the bugs}}$ based on a recommendation list with size $k$. We employ two typical classifiers, i.e., Naive Bayes (NB) and Supporting Vector Machine (SVM). These classifiers are implemented by Weka [33].

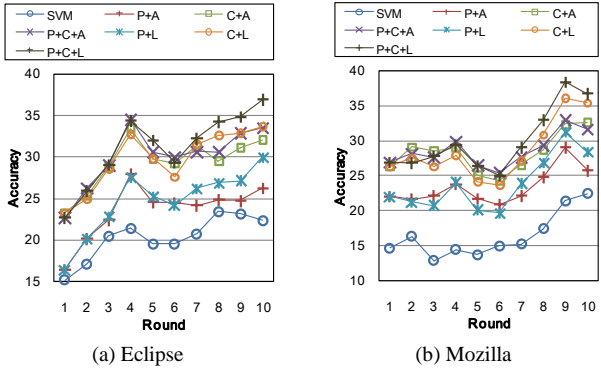

(a) Eclipse  (b) Mozilla

Figure 8. Accuracy for the top-1 predicted developer in 10-round incremental learning. SVM denotes the accuracy only based on the classifier, P and C denote the approach combining with the developer prioritization based on products and components, respectively. A and L denotes the approach based on the developer proritization, which is extracted from the accumulative folds and the latest fold in the training set, respectively.

In the following of this sub-section, we present the experimental results on bug triage. To obtain a high accuracy, we examine several ways to generate the developer prioritization from bug comments in training sets. We consider this problem on two dimensions, namely the source and the time period of developer prioritization. On one hand, in RQ1, we have presented that the developer prioritization changes in different products. Thus, we generate the developer prioritization based on products, components, and products & components, respectively. On the other hand, in RQ2, we have verified that the developer prioritization in two close periods can be very similar. Thus, we consider two kinds of time periods for building the developer prioritization, i.e., the accumulative folds in the training set and the latest fold in the training set.

Taken Round 2 in Fig. 7 as an example, Folds 1 and 2 are used to build the developer prioritization by choosing the accumulative folds while only the Fold 2 is used by choosing the last fold.

To improve bug triage, we combine the product-based and the component-based developer prioritization with the predicted results of classifiers. Given a new bug report, we extract its product and its component. We combine the probabilities in prediction with the developer priorities, which match its product and its component. Formally, given a new bug report $B$, its product $p$, and its component $c$, the final score is $F_i = P_i + (S_i^p + S_i^c)/M_s$, where $P_i$ is the probability predicted for each developer $d_i$ by a classifier, $S_i^p$ is the score in $p$-based developer prioritization, $S_i^c$ is the score in $c$-based developer prioritization, $M_s$ is the maximum value in $n$ developers for normalization ($M_s = max_{1 \leq i \leq n}(S_i^p + S_i^c)$). The score $S_i^p$ or $S_i^c$ is set to zero if a developer never appears in a product or a component. Then, we rank the developers by these final scores and select developers with highest scores as the final results. To obtain the rankings of top-$k$ developers for a bug report, we use a classifier to predict $2k$ developers and rank these developers by the combined final scores. Then, top-$k$ developers in the new rankings are selected as the final predicted results. In our work, we examine the results of top-5 developers.

In Fig. 8, we present the accuracy of the top-1 developer in each round. In both Eclipse and Mozilla, the approaches combined with the developer prioritization can obtain higher accuracy than directly using SVM. Most of the results based on the latest fold are better than those based on the accumulative folds. As a result, we observe that the recent changes of developers can be more helpful than the accumulative changes. This observation coincides with the existing work [1], [6] that the latest activity of a developer is representative. Among seven approaches in each sub-figure of Fig. 8, the approach combined with the developer prioritization based on products and components has the highest accuracy. In the rest of this sub-section, we choose such combination as the approach in the experiments.

TABLE IV PREFORMANCE OF BUG TRIAGE ON ECLIPSE AND MOZILLA

| Project | Classifier | Size | Approach | 1 | 2 | 3 | 4 | 5 | 6 | 7 | 8 | 9 | 10 | Average Accuracy | Improvement |
|---|---|---|---|---|---|---|---|---|---|---|---|---|---|---|---|
| Eclipse | SVM | Top 1 | SVM | 15.12 | 17.04 | 20.40 | 21.37 | 19.50 | 19.47 | 20.71 | 23.36 | 23.08 | 22.29 | 20.24 | 10.89 |
| | | | SVM+DP | 22.61 | 25.86 | 29.02 | 34.35 | 32.01 | 29.29 | 32.23 | 34.20 | 34.82 | 36.86 | 31.12 | |
| | | Top 3 | SVM | 27.41 | 30.70 | 34.62 | 38.20 | 33.60 | 34.02 | 37.93 | 40.27 | 39.16 | 40.48 | 35.64 | 12.94 |
| | | | SVM+DP | 38.35 | 42.29 | 46.84 | 52.30 | 47.02 | 45.51 | 52.06 | 53.51 | 52.27 | 55.60 | 48.57 | |
| | | Top 5 | SVM | 32.65 | 36.49 | 43.41 | 45.87 | 42.31 | 42.44 | 46.97 | 48.41 | 48.42 | 49.13 | 43.61 | 9.49 |
| | | | SVM+DP | 40.16 | 45.34 | 52.12 | 55.86 | 51.15 | 51.17 | 57.67 | 59.02 | 58.41 | 60.05 | 53.10 | |
| | NB | Top 1 | NB | 27.37 | 28.43 | 28.14 | 30.97 | 29.00 | 28.03 | 30.35 | 28.63 | 29.01 | 30.78 | 29.07 | 2.88 |
| | | | NB+DP | 30.57 | 32.12 | 31.28 | 33.47 | 32.07 | 30.55 | 32.96 | 31.43 | 31.77 | 33.23 | 31.94 | |
| | | Top 3 | NB | 33.73 | 35.41 | 34.13 | 36.32 | 34.33 | 33.05 | 36.16 | 34.84 | 34.27 | 36.15 | 34.84 | 1.88 |
| | | | NB+DP | 35.94 | 37.82 | 36.54 | 38.28 | 36.38 | 35.01 | 37.80 | 36.36 | 35.44 | 37.59 | 36.72 | |
| | | Top 5 | NB | 34.88 | 36.60 | 35.10 | 37.25 | 35.19 | 33.73 | 36.91 | 35.74 | 34.78 | 36.97 | 35.72 | 1.17 |
| | | | NB+DP | 36.16 | 38.00 | 36.58 | 38.40 | 36.63 | 35.12 | 37.93 | 36.43 | 35.66 | 37.92 | 36.88 | |
| Mozilla | SVM | Top 1 | SVM | 14.66 | 16.24 | 12.83 | 14.41 | 13.69 | 14.95 | 15.20 | 17.36 | 21.39 | 22.39 | 16.31 | 13.64 |
| | | | SVM+DP | 26.91 | 26.84 | 27.78 | 29.39 | 26.41 | 25.01 | 29.08 | 32.96 | 38.43 | 36.74 | 29.95 | |
| | | Top 3 | SVM | 29.93 | 37.30 | 34.78 | 33.92 | 34.85 | 33.24 | 36.56 | 39.97 | 44.54 | 45.40 | 37.05 | 12.87 |
| | | | SVM+DP | 42.33 | 48.87 | 45.60 | 47.75 | 44.70 | 43.98 | 50.29 | 55.46 | 61.04 | 59.13 | 49.91 | |
| | | Top 5 | SVM | 37.26 | 47.00 | 44.84 | 45.06 | 44.66 | 43.41 | 47.41 | 52.23 | 56.69 | 57.40 | 47.60 | 9.38 |
| | | | SVM+DP | 46.78 | 54.76 | 53.47 | 54.19 | 52.61 | 53.58 | 58.30 | 62.44 | 67.61 | 66.03 | 56.98 | |
| | NB | Top 1 | NB | 27.60 | 25.04 | 24.54 | 24.69 | 24.90 | 24.25 | 27.64 | 29.19 | 31.24 | 32.14 | 27.12 | 2.30 |
| | | | NB+DP | 30.08 | 27.56 | 26.59 | 26.88 | 26.70 | 26.52 | 29.91 | 31.13 | 34.15 | 34.76 | 29.43 | |
| | | Top 3 | NB | 33.78 | 32.20 | 30.65 | 30.08 | 31.55 | 31.76 | 34.69 | 35.73 | 37.78 | 39.25 | 33.75 | 2.36 |
| | | | NB+DP | 36.87 | 34.75 | 33.24 | 32.45 | 33.45 | 33.96 | 36.99 | 37.28 | 40.22 | 41.91 | 36.11 | |
| | | Top 5 | NB | 35.14 | 33.96 | 31.98 | 31.15 | 32.84 | 33.35 | 35.69 | 36.77 | 39.18 | 40.80 | 35.09 | 1.56 |
| | | | NB+DP | 37.15 | 34.96 | 33.74 | 32.81 | 34.32 | 34.93 | 37.53 | 37.81 | 40.83 | 42.38 | 36.65 | |

TABLE V  NUMBER OF BUGS FOR EACH SEVERITY

| Project | Product: Component | # Non-severe bugs | # Severe bugs |
|---|---|---|---|
| Eclipse | Platform: UI | 1425 | 3284 |
| | JDT: UI | 1425 | 1533 |
| | JDT: Text | 816 | 558 |
| Mozilla | Core: Layout | 255 | 929 |
| | Camino: Bookmarks | 62 | 40 |
| | Firefox: General | 2658 | 10124 |

TABLE VI  PERFORMANCE ON SEVERITY IDENTIFICATION

| Project | Product: Component | Non-severe bugs (%) | | | Severe bugs (%) | | |
|---|---|---|---|---|---|---|---|
| | | Precision | Recall | F-measure | Precision | Recall | F-measure |
| Eclipse | Platform: UI | 49.6 | 53.6 | 51.6 | 79.1 | **76.4** | 77.8 |
| | | 50.1 | **57.4** | **53.5** | 80.3 | 75.2 | 77.6 |
| | JDT: UI | 61.9 | 66.1 | 63.9 | 66.3 | 62.1 | 64.2 |
| | | 62.0 | 66.3 | 64.1 | 66.6 | 62.3 | 64.4 |
| | JDT: Text | 78.2 | 76.5 | 77.3 | 66.7 | 68.8 | 67.7 |
| | | 78.0 | 76.7 | 77.3 | 66.7 | 68.3 | 67.5 |
| Mozilla | Core: Layout | 45.5 | 65.1 | 53.5 | 89.1 | **78.6** | 83.5 |
| | | 45.6 | **71.4** | **55.7** | **90.7** | 76.6 | 83.1 |
| | Camino: Bookmarks | 76.6 | 79.0 | 77.8 | 65.8 | 62.5 | 64.1 |
| | | **77.6** | **83.9** | **80.6** | **71.4** | 62.5 | **66.7** |
| | Firefox: General | 42.8 | **50.6** | **46.4** | 86.4 | 82.2 | 84.3 |
| | | **46.2** | 36.4 | 40.7 | 84.2 | **88.9** | **86.5** |

In Table IV, we present the results of bug triage by combining the developer priorities with the output of the classifier in Eclipse and Mozilla. SVM+DP and NB+DP denote the results, which are ranked by combining the developer prioritization with the output of SVM and NB, respectively. We can find that for both of SVM and NB, the accuracy is improved when combining with the developer prioritization. The average improvement for SVM is around 10% while the average improvement for NB is only about 2%. The reason for such results is that the combination with the developer prioritization is based on the probability predicted by classifiers and SVM has stronger ability on discriminating different classes than NB. Thus, for NB, the ranks of developers may not change too much by combining the developer prioritization. As shown in Table IV, the improvement for the top 1 is larger than that for top 5. The improvement for SVM in some rounds are over 15%, e.g., Round 10 with top 3 developers in Eclipse and Round 9 with top 3 in Mozilla.

*B. Severity Identification*

Since the number of daily bugs is large to handle, new bugs in bug repositories are divided into different severities to process for different goals. Existing work (robotic severity [23], security severity [14], and critical severity [20]) has identified bug severities with predictive techniques.

In this sub-section, we address the critical severity problem. Lamkanfi et al. [20] predict whether a new bug is severe by adapting a text classification technique. In their work, a set of bug reports, including non-severe or severe ones, are divided into a training set and a test set. The 10-fold cross validation [33] is used to evaluate the results of classification. For a bug report with a self-reported severity, *trivial* or *minor*, is considered as a *non-severe* bug while *major*, *critical*, or *blocker* is considered as a *severe* one. Given a new bug report, the title of this bug is extracted to build the vector space model. The stop word removal, stemming techniques are performed (note that no *tf-idf* technique is used). Based on the vector space of training sets, Naive Bayes (NB) classifier is employed to identify whether a bug is severe or not.

In existing work by Lamkanfi et al. [20], the severity of a bug is predicted by the numeric vector, which is converted from the bug title. We consider adding the factors of developers to predict severities. For new bugs, the priorities of reporters may provide more information to identify the severity, e.g., a reporter with a high priority may pay more attentions to the severe bugs. Therefore, for each bug report, we add two features from the developer prioritization to the original numeric vectors. These two features are two priority scores of reporters based on the products and components, respectively. Then the predictive vector has two more numeric features. Since values in predictive vectors may be imbalance, we normalize all the features to the range from 0 to 1 for each training set.

In our work, we extract the data sets from bug repositories of Eclipse (bugs 1-300000) and Mozilla (bugs 300001-600000). For each project, we select *severe* and *non-severe* bugs in three components in accordance with [20]. These data sets are presented in Table V. We evaluate the severity identification with *precision*, *recall*, and *F-measure*.

We present the experimental results in Table VI. For each component, the first row denotes the classification based on the original vectors, which are extracted from bug titles, while the second row denotes the classification based on the mixed vectors, which are formed by adding two developer scores. A value in bold denotes a result, which is over 1% better than the other result on the same component.

As shown in Table VI, the prediction based on the mixed vectors can obtain better precision, recall, and F-measure on bugs in Mozilla while the difference of results in Eclipse is not significant. One possible reason for these results is the number of original numeric features is large and the effects of two new dimensions are not obvious to the predictive model.

*C. Reopened Bug Prediction*

Reopened bug prediction aims to identify a bug report, which will be incorrectly fixed in the future. To our knowledge, only Shihab et al. [30] have proposed the method to predict a reopened bug. In their work, they extract 22 factors in 4 dimensions to build the predictive model based on AdaBoost [33]. Among the 4 dimensions of factors, one dimension is called "people dimension", which consists of 4 factors about the developer information of a bug report, i.e., the reporter name, the fixer name, the reporter experience, and the fixer experience.

In this paper, we do not discuss the improvement of the predictive model of reopened bug prediction. We only con-sider whether our developer prioritization can provide similar factors to this model. We evaluate the experiments with the 10-fold cross validation on the benchmark data set in [30]. This data set consists of 1530 bugs (246 are reopened bugs and 1284 are not), which are extracted from bugs 1-300000 in Eclipse. We generate the developer prioritization (without specifying a product or a component) on the bug comments of bugs 1-300000 and extract the scores of reporters and fixers, who are specified in the data set. Note that we have not conducted experiments on Mozilla since only the benchmark for Eclipse is shared in [30].

TABLE VII FACTORS ON REOPENED BUG PREDICTION IN ECLIPSE

| Factor | Reopened bugs (%) | | | Non-reopened bugs (%) | | | Accuracy |
|---|---|---|---|---|---|---|---|
| | Precision | Recall | F-measure | Precision | Recall | F-measure | |
| (Original)Name+Exp. | 92.8 | 97.0 | 94.8 | 79.3 | 60.6 | 68.7 | 91.11 |
| Exp.+Priority | 93.1 | 96.5 | 94.8 | 77.4 | 62.6 | 69.2 | 91.05 |
| Name+Priority | 93.0 | 97.4 | 95.2 | 82.2 | 61.8 | 70.5 | 91.70 |
| Name+Exp.+Priority | 93.1 | 97.4 | 95.2 | 82.4 | 62.6 | 71.1 | 91.83 |

In the experiments, we add two factors to the people dimension, namely the reporter priority score and the fixer priority score. These 6 factors can be divided into three groups, i.e., *name*, *experience*, and *priority*. We list 4 combinations of these three groups of factors in the column "factor" in Table VII (Exp. is short for the group of factors *experience*). We run the algorithm AdaBoost in [30] to examine the results.

As shown in Table VII, by changing the original factors of *name* and *experience* with *priority*, the performance is improved. And the combination of all the three groups can provide the highest performance. We summarize the results in Table VII that the factors based on the developer prioritization can provide similar features to the model for predicting reopened bugs and can slightly improve the existing results. Note that the improvement by adding the factors of priority scores is not significant. This fact suffers from two possible reasons. One is the small number of training set may limit the predictive ability of factors; the other is the existing results may be good enough and hard to improve.

**Answer to RQ4**. By examining three typical tasks in bug repositories, we conclude that the developer prioritization is helpful to improve the results of these tasks, especially, the task of bug triage.

## VI. THREATS TO VALIDITY

### A. Building the Developer Prioritization

In our work, we build the developer prioritization from bug comments in bug repositories. The developer rankings are obtained to denote the priorities of developers. Since there is no real ranking of developers in software projects, it is hard to validate whether our obtained rankings are coincident with the real collaboration in development. To address this problem, a good way is to conduct a case study to examine the differences between the developer prioritization and the collaboration in development. Moreover, the developer prioritization in our work can enhance simple measures, e.g., distinguishing the developers with the same out-degree. An experiment should be conducted to compare the priorities between our work and simply measuring out-degrees.

Besides the developer prioritization for a whole project, we have also proposed product-based and component-based developer prioritization in our work. We extract the products and components from the self-reported terms of bug reports. However, such terms of products or components may mismatch the real terms since reporters lack experiences to identify the correct products or components [40]. To completely avoid such mismatch, a technique for identifying the correct product and components should be developed.

### B. Analysis of Developer Prioritization

In Section IV, we explore the evolution of developer prioritization over time. The time period is divided into several half years to study the changes in the process of development. Besides the time-based evolution, version-based evolution of the developer prioritization may provide more information. We do not investigate version-based evolution in our work since the bug reports in different versions are hard to collect. For example, among bugs 200001-300000, only 7450 bugs (7.45%) are identified with version information, which belong to 38 products.

To analyze the noise tolerance of developer prioritization, we recognize noisy comments with a heuristic, which views the comments by inactive developers and the comments with few words as noises. Many of such comments consist of noisy information, but some meaningful comments may also be viewed as noises, such as comments with only one word "fixed" or "verified". To exactly recognize noisy comments, manually labeling is more helpful than a heuristic.

### C. Assisting Software Tasks

In this paper, we show that the developer prioritization is effective to improve the tasks in bug repositories based on the empirical evaluation. The developer priorities can add more information to the input features or update the output of classifiers. However, further questions may be proposed, for example, what is the internal relationship between the developer prioritization and the goal of a task? And why is the social behavior of developers helpful? These questions are not easy to answer. In this paper, the developer prioritization can build a bridge from bug repositories to predictive tasks. For further work, a systematical case study can provide more information to explore the correlation between the developer prioritization and the predictive tasks.

## VII. RELATED WORK

### A. Social Network Analysis in Software Repositories

Bird et al. [3] mine social networks from email achieves and analyze the developer activity based on social network measures. Their later work [8] explores communities from the social networks, which are representative of the collaboration of developer behavior. Wolf et al. [34] employ the team communication network to predict the failures of software builds. In their work, the centrality measures in social networks are extracted as features in the predictive model.

Social networks have been proposed to improve the software failure prediction. To predict failures, Meneely et al. [25] construct developer networks on change log repositories and Pinzger et al. [27] build developer-module networks on binary repositories. Bird et al. [7] combine the topological properties with social networks and investigate multiple types of relationships to predict failures. Moreover, Bettenburg & Hassan [4] study the impacts of socio-technical measures through the failure detection. Since the social techniques are effective for indicating software failures, Meneely & Williams [24] empirically validate that the social network metrics can represent the collaboration relationship in software development.

By building a social networking service for developers, Begel et al. [5] introduce a Codebook framework to discover the inter-team coordination in development. They have conducted two applications to evaluate the effectiveness of their framework.

In bug repositories, Hong et al. [17] have examined the developer social networks in bug repositories. They discover the sub-communities of a developer network and investigate the evolution over time. In this paper, we also focus on the socio-technical analysis on bug repositories. In contrast to the community discovery in [17], we explore the developer prioritization in bug repositories. We analyze the developer rankings and improve three existing tasks in bug repositories.

*B. Bug Repositories*

In this paper, we investigate the developer prioritization in bug repositories. In existing work on bug repositories, Fischer et al. [13] explore the proximity of software features. In their work, bug report analysis is used to study and visualize the relations between software features.

The quality of bug reports is important for locating and fixing bugs. Hooimeijer & Weimer [18] present the first work to model the quality of bug reports. Bettenburg et al. [9] point out that duplicate bugs contain extra information for bug fixing, which are useful to improve bug triage. Zimmermann et al. [40] study the evidence for the mismatch between developer expectation and bug reports based on a systematic questionnaire survey. Xiao & Afzal [36] propose a search-based approach for the resource scheduling on bug fixing tasks.

Most work on bug repositories treats developers and bugs separately. Our previous work [37] proposes an integrated view of developers and bug reports. We transfer the interactions between developers and bug reports to requirements engineering to supplement the lack of open requirements.

In this paper, we extract the developer prioritization in bug repositories. Besides the multiple aspects of analysis, we leverage the developer prioritization to predict software tasks.

*C. Predicting Tasks in Bug Repositories*

The goal of bug triage is to automatically assign a new bug to the correct developer to avoid the expensive cost of maintenance. Čubranić & Murphy [10] have proposed the first work of bug triage, which transforms bug triage to a text categorization problem. Anvik et al. [1] extend the above work with a recommendation list and multiple classifiers. Jeong et al. [19] and Bhattacharya & Neamtiu [6] propose a tossing graph based approach to improve bug triage with the previous assignment history of bug reports. Our previous work [38] proposes a semi-supervised learning approach to avoid the lack of qualified bug reports. Recent work by Anvik & Murphy [2] investigates the effects of recommenders to assist bug triage for streamlining the development process. Other work also addresses the problem of bug triage, such as the training set reduction [39], the fuzzy-set and cache-based approach [32], and the cost-aware bug triage [26].

Severity identification is to detect the bug severities to guide the resource allocation and planning of bug fixing. To date, three types of severities are studied. Menzies & Macus [23] first propose a text mining approach to detect the 5-level robotic severity for bug reports in NASA databases. Gegick et al. [14] and Lamkanfi et al. [20] have further predicted the security severity and the critical severity of bug reports.

Reopened bug prediction is to detect whether a bug is fixed in a correct way. Shihab et al. [30] study and predict reopened bugs on 22 factors in four dimensions, which are extracted from both bug repositories and source code repositories. In an empirical study on characterizing which bugs get fixed, Guo et al. [15] present that the times of reopenings is a factor to indicate whether a bug can be fixed.

In this paper, we empirically evaluate whether the developer prioritization can improve the results of the above three tasks. Besides the three mentioned tasks, existing work improves the software quality in bug repositories on other tasks. For example, Rastkar et al. [29] summarize long bug reports to avoid redundancies and noises in bug repositories; Runeson et al. [28], Wang et al. [35], and Sun et al. [31] detect duplicate bug reports to reduce the expense for handling bugs in large scale repositories.

## VIII. CONCLUSION AND FUTURE WORK

In this paper, we model the developer prioritization in bug repositories by extending a socio-technical approach. We analyze three problems of the developer prioritization, namely the characteristics in products, the evolution, and the tolerance of noises. Based on the analysis, we investigate the ways to leverage the developer prioritization to improve three typical tasks in bug repositories. The results are studied on over 900000 bug reports in Eclipse and Mozilla.

Our future work is to investigate a task-based developer prioritization in bug repositories to improve a specified task with the developer rankings. In contrast to the general model of the developer prioritization, we want to provide a model to add more knowledge to handle the problems in a specified task. For example, fixers of bug reports should be added more weights in the developer prioritization to improve bug triage.

ACKNOWLEDGMENT

We greatly thank the anonymous reviewers for their insightful comments. This work is partially supported by the National Natural Science Foundation of China under grants

61175062, 60805024, and 61033012, and the "Software + X" funding of Dalian University of Technology.

REFERENCES

[1] J. Anvik, L. Hiew, and G.C. Murphy, "Who Should Fix This Bug?," Proc. 28th Intl. Conf. Software Engineering (ICSE '06), May 2006, pp. 361-370.

[2] J. Anvik and G.C. Murphy, "Reducing the Effort of Bug Report Triage: Recommenders for Development-Oriented Decisions," ACM Trans. Software Engineering & Methodology, vol.20, no.3, Aug. 2011.

[3] C. Bird, A. Gourley, P. Devanbu, M. Gertz and A. Swaminathan, "Mining Email Social Networks," Proc. 3rd Intl. Workshop Mining Software Repositories (MSR '06), May 2006, pp.137-143.

[4] N. Bettenburg and A.E. Hassan, "Studying the Impact of Social Structures on Software Quality," Proc. IEEE 18th Intl. Conf. Program Comprehension (ICPC '10), Jun. 2010, pp.124-133.

[5] A. Begel, Y.P. Khoo, and T. Zimmermann, "Codebook: Discovering and Exploiting Relationships in Software Repositories," Proc. 32nd Intl. Conf. Software Engineering (ICSE '10), May 2010, pp. 125-134.

[6] P. Bhattacharya and I. Neamtiu, "Fine-Grained Incremental Learning and Multi-Feature Tossing Graphs to Improve Bug Triaging," Proc. 26th IEEE Intl. Conf. Software Maintenance (ICSM '10), Sept. 2010, pp. 1-10.

[7] C. Bird, N. Nagappan, H. Gall, B. Murphy, and P. Devanbu, "Putting it All Together: Using Socio-Technical Networks to Predict Failures," Proc. 20th Intl. Symp. Software Reliability Engineering (ISSRE '09), Nov. 2009, pp.109-119.

[8] C. Bird, D. Pattison, R. D'Souza, V. Filkov and P. Devanbu, "Latent Social Structure in Open Source Projects," Proc. 16th ACM SIGSOFT Intl. Symp. Foundations of software engineering (FSE '08), Nov. 2008, pp.24-35.

[9] N. Bettenburg, R. Premraj, T. Zimmermann, and S. Kim, "Duplicate Bug Reports Considered Harmful… Really?," Proc. 24th IEEE Intl. Conf. Software Maintenance (ICSM '08), Sept. 2008, pp. 337-345.

[10] D. Čubranić and G.C. Murphy, "Automatic Bug Triage Using Text Categorization," Proc. 16th Intl. Conf. Software Engineering & Knowledge Engineering (SEKE '04), Jun. 2004, pp. 92-97.

[11] T. DeMarco and T. Lister, Peopleware: Productive Projects and Teams, 2nd ed. Dorset House, New York, 1999.

[12] Eclipse. http://eclipse.org/.

[13] M. Fischer, M. Pinzger and H. Gall, "Analyzing and Relating Bug Report Data for Feature Tracking," Proc. 10th Working Conf. Reverse Engineering (WCRE '03), Nov. 2003, pp. 90-101.

[14] M. Gegick, P. Rotella, and T. Xie, "Identifying Security Bug Reports via Text Mining: An Industrial Case Study," Proc.7th IEEE Working Conf. Mining Software Repositories (MSR '10), May 2010, pp. 11-20.

[15] P.J. Guo, T. Zimmermann, N. Nagappan, and B. Murphy, "Characterizing and Predicting Which Bugs Get Fixed: An Empirical Study of Microsoft Windows," Proc. 32nd Intl. Conf. Software Engineering (ICSE '10), May 2010, pp. 495-504.

[16] A.E. Hassan, "The Road Ahead for Mining Software Repositories," Proc. Frontiers of Software Maintenance (FoSM '08), Sept. 2008, pp. 48-57.

[17] Q. Hong, S. Kim, S.C. Cheung, and C. Bird, "Understanding a Developer Social Network and its Evolution," Proc. 27th IEEE Intl. Conf. Software Maintenance (ICSM '11), Sept. 2011, pp. 323-332.

[18] P. Hooimeijer and W. Weimer, "Modeling Bug Report Quality," Proc. 22nd IEEE/ACM Intl. Conf. Automated Software Engineering (ASE '07), Nov. 2007, pp. 34-43.

[19] G. Jeong, S. Kim, and T. Zimmermann, "Improving Bug Triage with Tossing Graphs," Proc. 17th ACM SIGSOFT Symp. Foundations of Software Engineering (FSE '09), Aug. 2009, pp. 111-120.

[20] A. Lamkanfi, S. Demeyer, E. Giger, and B. Goethals, "Predicting the Severity of a Reported Bug," Proc. 7th IEEE Working Conf. Mining Software Repositories (MSR '10), May 2010, pp. 1-10.

[21] L. Lü, Y.-C. Zhang, C.H. Yeung, and T. Zhou, "Leaders in Social Networks, the Delicious Case," PLoS One, vol. 6, no. 6, Jun. 2011.

[22] Mozilla. http://mozilla.org/.

[23] T. Menzies and A. Marcus, "Automated Severity Assessment of Software Defect Reports," Proc. IEEE Conf. Software Maintenance (ICSM '08), Sept. 2008, pp. 346-355.

[24] A. Meneely and L. Williams, "Socio-Technical Developer Networks: Should We Trust Our Measurements?," Proc. 33rd Intl. Conf. Software Engineering (ICSE '11), May 2011, pp. 281-290.

[25] A. Meneely, L. Williams, W. Snipes, and J. Osborne, "Predicting Failures with Deverloper Networks and Social Network Analysis," Proc. 16th ACM SIGSOFT Intl. Symp. Foundations of Software Engineering (FSE '08), Nov. 2008, pp. 13-23.

[26] J.-W. Park, M.-W. Lee, J. Kim, S.-W. Hwang, and S. Kim, "CosTriage: A Cost-Aware Triage Algorithm for Bug Reporting Systems," Proc. 25th Conf. Artificial Intelligence (AAAI '11), Aug. 2011, pp. 139-144.

[27] M. Pinzger, N. Nagappan, and B. Murphy, "Can Developer-Module Networks Predict Failures?," Proc. 16th ACM SIGSOFT Intl. Symp. Foundations of Software Engineering (FSE '08), Nov. 2008, pp. 2-12.

[28] P. Runeson, M. Alexandersson, and O. Nyholm, "Detection of Duplicate Defect Reports Using Natural Language Processing," Proc. Intl. Conf. Software Engineering (ICSE '07), May 2007, pp. 499-510.

[29] S. Rastkar, G.C. Murphy, and G. Murray, "Summarizing Software Artifacts: A Case Study of Bug Reports," Proc. 32nd Intl. Conf. Software Engineering (ICSE '10), May 2010, pp. 505-514.

[30] E. Shihab, A. Ihara, Y. Kamei, W.M. Ibrahim, M. Ohira, B. Adams, A.E. Hassan, and K. Matsumoto, "Predicting Re-opened Bugs: A Case Study on the Eclipse Project," Proc. 17th Working Conf. Reverse Engineering (WCRE '10), Oct. 2010, pp. 249-258.

[31] C. Sun, D. Lo, X. Wang, J. Jiang, and S.-C. Khoo, "A Discriminative Model Approach for Accurate Duplicate Bug Report Retrieval," Proc. 32nd Intl. Conf. Software Engineering (ICSE '10), May 2010, pp. 45-54.

[32] A. Tamrawi, T.T. Nguyen, J.M. Al-Kofahi, and T.N. Nguyen, "Fuzzy-Set and Cache-Based Approach for Bug Triaging," Proc. 19th ACM SIGSOFT Symp. Foundations of Software Engineering (FSE '11), Sept. 2011, pp. 365-375.

[33] I.H. Witten, E. Frank, and M.A. Hall, Data Mining: Practical Machine Learning Tools and Techniques, 3rd ed. Morgan Kaufmann, Burlington, MA, 2011.

[34] T. Wolf, A. Schröter, D. Damian, and T. Nguyen, "Predicting Build Failures Using Social Network Analysis on Developer Communication," Proc. 31st Intl. Conf. Software Engineering (ICSE '09), May 2009, pp. 1-11.

[35] X. Wang, L. Zhang, T. Xie, J. Anvik, and J. Sun, "An Approach to Detecting Duplicate Bug Reports Using Natural Language and Execution Information," Proc. 30th Intl. Conf. Software Engineering (ICSE '08), May 2008, pp. 461-470.

[36] J. Xiao and W. Afzal, "Search-based Resource Scheduling for Bug Fixing Tasks," Proc. 2nd Intl. Symp. Search Based Software Engineering (SSBSE '10), Sept. 2010, pp. 133-142.

[37] J. Xuan, H. Jiang, Z. Ren, Z. Luo, "Solving the Large Scale Next Release Problem with a Backbone Based Multilevel Algorithm," IEEE Trans. Software Engineering, preprint, doi: 10.1109/TSE.2011.92.

[38] J. Xuan, H. Jiang, Z. Ren, J. Yan, and Z. Luo, "Automatic Bug Triage Using Semi-Supervised Text Classification," Proc. 22th Intl. Conf. Software Engineering & Knowledge Engineering (SEKE '10), Jul. 2010, pp. 209-214.

[39] W. Zou, Y. Hu, J. Xuan, and H. Jiang. "Towards Training Set Reduction for Bug Triage," Proc. 35th Annual IEEE Intl. Computer Software and Applications Conference (COMPSAC '11), Jul. 2011, pp. 576-581.

[40] T. Zimmermann, R. Premraj, N. Bettenburg, S. Just, A. Schröter, and C. Weiss, "What Makes a Good Bug Report?," IEEE Trans. Software Engineering, vol. 36, no.5, Oct. 2010, pp. 618-643.